\documentclass[letter]{aa} 

\usepackage{graphicx}
\usepackage{txfonts}
\usepackage{color}

\newcommand{\msol}{M$_{\odot}$}
\newcommand{\hmass}{$h^{-1}$M$_{\odot}$}
\newcommand{\hmpc}{$h^{-1}$Mpc}
\newcommand{\hunit}{km\,s$^{-1}$\,Mpc$^{-1}$}
\newcommand{\kms}{km\,s$^{-1}$}
\newcommand{\hconst}{H$_{0}$}

\begin{document} 

\title{On the definition of superclusters}
\author{Gayoung Chon\inst{1}, Hans B\"ohringer\inst{1}, Saleem Zaroubi\inst{2}
}
\institute{Max-Planck-Institut f\"ur extraterrestrische Physik,
  Giessenbachstrasse, Garching D-85748 Germany
  \and 
  Kapteyn Astronomical Institute, University of Groningen,
  PO Box 800, 9700 AV Groningen, the Netherlands\\
  \email{gchon@mpe.mpg.de}
}

\date{Received 28 12, 2014; accepted 06 02, 2015}

\abstract
{
To obtain a physically well-motivated definition of superclusters, we proposed in our previous work
to select superclusters with an overdensity criterion that selects only those objects that will collapse in 
the future, including those that are at a turn-around in the present epoch.  
In this paper we present numerical values for these criteria for a range of standard cosmological models.
We express these criteria in terms of a density ratio or, alternatively, as an infall velocity and 
show that these two criteria give almost identical results. 
To better illustrate the implications of this definition, we applied our criteria to some prominent structures
in the local Universe, the Local supercluster, Shapley supercluster, and the recently reported
Laniakea supercluster to understand their future evolution.
We find that for the Local and Shapley superclusters, only the central regions will collapse in the future,
while Laniakea does not constitute a significant overdensity and will disperse in the future. 
Finally, we suggest that those superclusters that will survive the accelerating cosmic expansion
and collapse in the future be called ``superstes-clusters'', where ``superstes'' means survivor in Latin, 
to distinguish them from traditional superclusters.
}


\maketitle

\section{Introduction}

Superclusters are the largest prominent density enhancements in our Universe. 
In the framework of hierarchical structure formation, superclusters are the next objects up from clusters, 
but unlike clusters, they are not virialised. 
They are generally defined as groups of two or more galaxy clusters above a certain spatial density 
enhancement~\citep{bahcall-88}. 
In this sense superclusters have mostly been treated just as a collection of clusters. 
Without a clear definition, we are left with structures with heterogeneous properties. 
Unlike clusters, superclusters have not reached a quasi-equilibrium configuration that defines
their structure.
As we observe them today, they are transition objects that largely reflect their initial conditions.
In contrast, clusters can be approximately described by their equilibrium configuration,
as given by, for example, the NFW model~\citep{navarro95}. 
Even disturbed and merging clusters are characterised by their deviations from this model.

For transition objects like superclusters, such a description is not possible.
One solution to this problem is to include the future evolution in the definition of the object, 
selecting only those superclusters that will collapse in the future in a more homogeneous class of objects.
There have been attempts to explore this definition in simulations~\citep{duenner06}. 
We have been exploring a similar approach observationally in our construction of an X-ray supercluster 
catalogue~\citep{chon13,chon14}. 
We selected the superclusters from the X-ray galaxy cluster distribution by means of a friends-of-friends 
(fof) algorithm in such a way that we expect the superclusters to have their major parts gravitationally 
bound and to collapse in the future. 
We found that we obtain a good understanding of the properties of our supercluster sample, and 
we can recover many of the known superclusters described in the literature in our survey volume. 

This selection is found to be slightly more conservative by not linking all the surrounding structure to 
the superclusters, which are linked to these objects in some other works. 
In other cases, such as for the Shapley supercluster, large structures are split into substructures. 
But overall the sample has good properties. 
Thus we find our method not only physically well motivated, but also appealing in selecting the structures 
that appear observationally distinct and prominent. 
To distinguish objects selected by our definition from the general usage of the word
superclusters, we suggest that those systems be called ``superstes-clusters'', relating to the Latin word 
superstes, which means survivor. 
Considering origins, superclusters were first studied at the time when most cosmologists favoured
a marginally closed Universe in which all overdense regions would eventually collapse. 
It came with the general acceptance of a re-accelerating Universe that this concept of future collapse
needed to be revised. We prefer using a new term over redefining the word supercluster, 
out of respect for the previous studies that were done with a less strict definition. 

It is our goal with this paper to explore this definition of superclusters and its consequences
in some more detail.
In particular, we provide numerical values for the selection criteria for various cosmologies. 
So far, we have based our selection criterion on the matter overdensity, which is motivated by our X-ray 
cluster observations. 
In theory, on large scales where the dynamics is dominated by gravity, observations of velocity fields
should closely reflect the dynamical evolution of structures and the underlying mass distribution. 
The velocity field also does not suffer from the bias that clusters and, to an extent, galaxies have. 
As a result, the dynamical information may provide a better basis for predicting the future 
evolution of the large-scale structure~\citep{zeldovich70,shandarin89,dekel94,zaroubi99}. 
Therefore we also explore the selection criteria in terms of the infall velocity.  
In our case we expect a close correspondence between the overdensity and the streaming motions, since the 
large-scale structure at the scale of superclusters is still in the quasi-linear regime of structure formation. 
Nevertheless, we test the correspondence between the two descriptions in the paper. 

We lay out our concept and explain the criterion in Sec. 2. In Sec. 3 we illustrate the concept by
three applications and Sec. 4 provides discussion and summary.

\section{Theoretical concept}

To study which primordial overdensities will finally collapse, we approximate the overdense regions by spheres 
with homogeneous density. 
This approximation has been successfully used for many similar investigations.
We can then model the evolution of the overdensity with respect to the expansion of the background cosmology 
with reference to Birkhoff's theorem. 
This allows us to describe the evolution of both regions, overdensity and background, by the respective values of 
the local and global parameters of Hubble constant, H, matter density, $\Omega_m$, and the parameter 
corresponding to the cosmological constant, $\Omega_{\Lambda}$. 

We calculate the evolution of the local and the global regions by integrating the Friedmann equation 
for their dynamical evolution.
We use $z$ = 500 as the starting point of our integration, since we have found that with this starting redshift, 
the final results have an accuracy well below 1\%. 
To find the starting values, we find the relevant cosmological parameters mentioned above for the background 
cosmology at $z$ = 500.
We then define an overdense region by increasing $\Omega_m$ so that we find a collapsed region in the future, 
which is solved iteratively. 

In comparison with the structures seen today, we are interested in the following properties of these 
marginally collapsing objects, which should be observable. 
What are their typical matter overdensities in the current epoch? 
What is the Hubble parameter that characterises their current, local dynamical evolution? 
These parameters will depend on the characteristics of the background cosmology. 
We have therefore calculated the density and expansion parameters for collapsing overdensities in a set of 
relevant cosmologies. 
We assumed a flat cosmology, with $\Omega_m + \Omega_{\Lambda} = 1$ for most models with the exception
of one example for an open cosmology. 
Amongst the models shown are the best-fit results from the PLANCK~\citep{planckcosm} and WMAP~\citep{wmap9} 
missions. 

The parameters shown in Table~\ref{tab:condition} are the density ratio of the overdensity to the background density, 
$R = \rho_{\rm ov} / \rho_m$, the parameter characterising the homogeneous expansion of the overdense sphere 
given in form of a Hubble parameter, H$_{\rm ov}$, and the overdensity of the local region with respect to the 
critical density, $\rho_c$, of the Universe, $\Delta_c = (\rho_{\rm ov} - \rho_c)/ \rho_c$.
The results show that while the density ratio varies with the matter density in the background universe, 
the overdensity parameter with respect to critical density hardly changes. 
Changing the Hubble constant does not alter the nature of the solution. 
The parameters, $R$ and $\Delta_c$, stay constant, while the local expansion parameter, H$_{\rm ov}$, only scales 
with the Hubble parameter of the background cosmology. 
Therefore we show only one example for the change in the parameters with the Hubble constant. 

Another interesting characterisation for superclusters are those structures that are at turn-around now. 
These structures have decoupled from the Hubble flow already and are at rest in the Eulerian reference frame 
so are just starting to collapse now. 
Here the local Hubble parameter is zero by definition.
The Einstein--de Sitter(EdS) model with the parameter, $ \Omega_m = 1.0$, yields a value of  
$\Delta_c = (3\pi / 4)^2 -1 \sim 4.55$, which can also be calculated analytically.

\subsection{Comparison of overdensity and dynamical criteria}

The threshold parameters for collapse given in Table~\ref{tab:condition} are only valid for spherically symmetric overdensities,
which are reasonable approximations for realistic overdensities. 
As described in the previous section, the velocity field provides a better basis for predicting the 
future evolution of a large-scale structure than does the density distribution~\citep{dekel94,zaroubi99}. 
Also in observations, the density distribution of objects has to be corrected for their large-scale 
structure bias, which is not necessary for evaluating the velocity field. 
However, in current astronomical observations, peculiar velocity data are only available for the very local 
region of the Universe, and for most other applications, we only have estimates of overdensity. 
Therefore it is important to test how well our criteria that are based on the overdensity argument correspond 
to those on the velocity information for realistic supercluster morphologies. 

\begin{figure}[h]
  \centering
  \resizebox{\hsize}{!}{\includegraphics{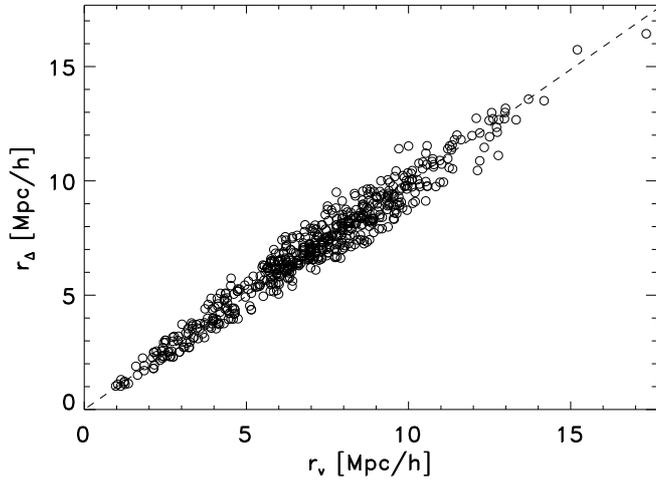}}
  \caption{
    Comparison of the two radii that define the collapsing region of a supercluster. 
    $r_{\Delta}$ is determined by the required density ratio criteria, and 
    $r_{\rm v}$ by the infall velocity criteria. 
    The dashed line indicates the one-to-one line. 
  }
  \label{fig:radii}
\end{figure}

For this reason we used the cosmological N-body simulations~\citep{springel05} to compare the radius 
inside which the structure is predicted to collapse based on the overdensity, $r_{\Delta}$, to that of the 
infall velocity, $r_{\rm v}$ for 570 superstes-clusters. 
For the details of the construction of the superclusters in simulations, we refer the readers to~\cite{chon14}. 
The value of $r_{\Delta}$ was taken at the radius where the density ratio reaches the threshold value
for collapse, and $r_{\rm v}$ is defined as the radius where the required infall velocity is reached. 
The latter radius marks the largest distance within which, on average, the infall velocities of  
all haloes are detached from the Hubble flow with the local expansion parameter prescribed in 
the previous section. 
The very close correspondence between the two predictions are shown in Fig.~\ref{fig:radii}.

We only have five pathological cases, where the collapse overdensity is only reached once away from the centre, 
while the infall pattern never reaches the required threshold. 
These are the cases where the most massive structures are concentrated not at the centre of the supercluster, 
but near the radius, $r_{v}$, and beyond.
In these cases the velocity pattern is very different from a smooth radial infall, and the supercluster 
is most probably fragmented into two or more massive substructures near the supercluster boundary in the future. 
We find it as a very strong encouragement for our approach, where the two alternative criteria 
usually give very similar results. 
The good correspondence is also a confirmation that structure evolution on the scale of superclusters is still 
in the quasi-linear regime.

\section{Applications}

In this section we illustrate the implications of our superstes-cluster definition with respect 
to some of the known superclusters.
The homogeneous sphere approximation only gives a rough estimate of the collapse situation. 
More detailed solutions have to take the morphology and substructure of the systems into account, 
which has actually been done for some of the cases below. 
Both criteria listed in Table~\ref{tab:condition}, the overdensity and the peculiar infall velocity, can be used 
for such a first estimate.
The aim of the following discussion is therefore only an approximate application of the suggested 
criteria for illustration. 
We use parameters for the cosmological model with $\Omega_m$=0.3 and \hconst=70~\hunit  
for the considerations below.

\subsection{Local supercluster and Virgo infall}

The Local supercluster is a high concentration of matter roughly centred on the Virgo cluster, 
which includes the Milky Way and the Local Group in the outskirts. 
It was first described by~\cite{dev53,dev58}. 
Detailed studies found the system to be mostly concentrated in an elongated filament that extends about 40~\hmpc, e.g.~\citet{tully78},~\citet{kara96},~\citet{lahav00},~\citet{klypin03}.
The Local Group is at a distance of about 16 to 17~Mpc from Virgo, and it shows a peculiar infall velocity
towards the Virgo cluster of about $220 \pm 30$~km~s$^{-1}$~\citep{sandage10} or 
$\sim 250$~km~s$^{-1}$~\citep{klypin03}.  

Applying the peculiar velocity criterion for future collapse, we find the following. 
The necessary peculiar infall velocity at a certain distance from the centre of the supercluster is 
given by the difference of the local to the background Hubble parameters multiplied by the distance, 
in our case $ d \times 49.5$~\hunit. 
Thus at the distance of the Local Group, a peculiar infall velocity of about 800~km~s$^{-1}$ would be required. 
Therefore the Local Group will recede from the Local supercluster in the distant future. 
This has been concluded in several previous works, and it has been shown, for example, by N-body simulations
by~\cite{nagamine03} based on a numerical constraint reconstruction of the local Universe by~\cite{mathis02}. 
Inspecting the velocity flow patterns of the Local supercluster shown in ~\citet{klypin03},~\citet{tully04},
and ~\citet{courtois12}, we find that only inside 10~Mpc infall velocities up to 500~km~s$^{-1}$ seem to occur, 
reaching the lower limit for a collapse. 
We can also use the estimate of the infall velocity profile from the constraint reconstruction of 
the Local supercluster by~\cite{klypin03} to find the outermost collapsing shell of the supercluster. 
They approximate the mean infall pattern by v = 145~(13~h$^{-1} {\rm Mpc}/r)^{1/2}$~km~s$^{-1}$.
With this prescription we find that only the regions inside about 5.5~Mpc will collapse in the future.   

We can also use the spherical collapse model to obtain a mass estimate of the Virgo cluster and
its surroundings from the peculiar velocity of the Milky way towards the Virgo cluster of $\sim$250~\kms.
When assuming a Virgo distance of 16~Mpc~\citep{tonry91}, the infall peculiar velocity corresponds 
to a local Hubble parameter of $\sim$54.5~\hunit for a homogeneous sphere with its centre at the Virgo distance. 
An integration of the Friedmann equations for our fiducial cosmology infers a ratio of the local 
overdensity to the cosmic mean of about 2.6.
This translates into a mass of Virgo and surroundings inside a radius of 16~Mpc of $1.8\times10^{15}$~\msol. 
\cite{tully04} get a mass of 1.2$\times10^{15}$~\msol\ from fitting the infall pattern, which probably gives 
most weight to the measurement at a slightly smaller radius.
\cite{klypin03} obtain a mass of $\sim10^{15}$~\msol\ for Virgo and the central filament of the local
supercluster from a constrained reconstruction of the local supercluster.
~\cite{kara14} studied the infall pattern of tracers with good distance estimates from the HST observations,
and find a mass of $8\times10^{14}$~\msol\ inside a radius of 7.2~Mpc which they estimate to be the
turn-around radius at the current epoch.
The fair agreement of the different methods shows that the spherical infall models provide
an excellent first estimate of the fate of such a supercluster.

\subsection{Laniakea}

Based on the compilation of peculiar velocities of galaxies out to $z$=0.1 in~\cite{tully13}, \cite{tully14} 
presented a new supercluster, which they call Laniakea. 
We refer the readers to~\cite{tully13} for the detailed methodology, but in essence they rely on 
the absolute distance measures estimated from six methods including the Tully-Fisher relation 
to calculate peculiar velocities of galaxies within 400~\hmpc, where the coverage is sparse beyond 100~\hmpc. 
In total they report distance measures for more than 8000 galaxies in the whole survey region. 
To reconstruct the underlying velocity field, they used the Wiener filter algorithm~\citep{zaroubi95}. 
They conclude from the reconstructed velocity field that there is a coherent flow within a sphere of 
radius, 80~\hmpc, which contains an estimated mass of $10^{17}$\hmass, and they define this region as a supercluster. 
Their mass estimate implies that the ratio of Laniakea density to the mean density is about 0.94.
Thus Laniakea does not even constitute a region with a significant overdensity and does not fulfil our
criteria in Table~\ref{tab:condition} for a supercluster.
As a result, Laniakea is far from being a bound system, and as a whole, it will disperse in the future, 
while only several dense regions will collapse within.

We can also look at Laniakea from another point of view by applying the peculiar velocity argument. 
If Laniakea was a bound structure with a radius of 80~\hmpc, we would estimate the required infall 
velocity at the boundary to be about 5700~\kms\ based on the local Hubble parameter estimate given in 
Table~\ref{tab:condition} for future collapse.  
However, typical peculiar velocities in their catalogue hardly exceed 500-700~\kms\ in the survey. 
This emphasises again that Laniakea as a whole is not a bound structure.

We also note that their value of the Hubble constant obtained by minimising the velocity monopole is 75.2$\pm$3.0~\hunit. 
There are also other local measurements of \hconst, notably by~\cite{riess11} based on 253 Type Ia supernovae 
data from the HST and by~\cite{freedman12} based on an additional mid-infrared observation of Cepheids. 
The former gives \hconst\ of 74.8$\pm$3.1, and the latter 74.3$\pm$2.1~\hunit, both with less than 3\% uncertainty.
Within their respective errors, the three measurements agree, which is also pointed out by~\cite{tully14}. 
It is interesting to compare these values to \hconst\ measured by the CMB experiments, WMAP, and Planck, 
where \hconst\ is more sensitive to very large scales. 
The best-fit \hconst\ values are 70.0$\pm$2.2~\citep{wmap9} and 68.0$\pm$1.4~\hunit\ \citep{planckcosm}, so local measurements of \hconst\ are always greater than those from the CMB measurements, although they would agree
within their respective current 1$\sigma$ uncertainty. 
With this it is interesting to note that \cite{turner92} pointed out that a locally underdense Universe would yield 
a Hubble constant larger than the cosmic mean. 
In reference to our work~\citep{r2density}, we find a local region within a radius of 170~\hmpc\  in the southern sky,
which is under-dense by $\approx$20\% with respect to the mean density resulting from the cluster number density 
that is under-dense by 40\% with a cluster bias of about two. 
Since the REFLEX survey has a high completeness up to a redshift of $z\sim$0.3, we have a good handle for tracing the 
matter density out to a very large radius, containing the volume of Laniakea. 
The amplitude of the local under-density traced by REFLEX II of $\approx$20\%, which implies that the locally 
measured \hconst\ would be 3$\pm$1.5\% more than for the cosmic one. 
It therefore indicates that the \hconst\ value obtained by~\cite{tully14} is consistent with the case where \hconst\ 
is measured in an under-dense region.

\subsection{Shapley supercluster}

The Shapley supercluster is known to have the highest concentration of galaxies in the nearby Universe,
at a redshift around $0.046$~\citep{scaramella89,raychaudhury89}. 
X-ray emission in the central region of Shapley was first mapped by~\cite{kull99}, and it
even traces the intra-cluster gas. 
We also report three X-ray superclusters in the area of the Shapley supercluster, constructed with 
a cluster overdensity parameter, $f$=10, from the REFLEX II cluster catalogue~\citep{chon13}. Here,  
$f$ is related to the density ratio $R$ by, $f=(R-1)b_{\rm CL}+1$ where $b_{\rm CL}$ is a cluster bias factor. Even with a generous choice of $f$=10, 
this already indicates that Shapley will be broken into 
smaller concentrations of matter, rather than collapsing into one large object. 
To get deeper insight, we calculated density ratios with data taken from the literature and from our X-ray 
work described below. 

\cite{reisenegger00} used velocity caustics from 3000 galaxies to define a central region of Shapley
and estimated the upper bound mass in a spherical radius of 8~\hmpc\ to be 1.3$\times10^{16}$\hmass. 
In this case the density ratio, $R,$ is 20.7. 
\cite{ragone06} established a mass function of about 180 systems with redshifts in the Shapley region
and give a lower mass limit of 1.1$\times10^{16}$\hmass within the same volume as above.
In this case, $R$ is about 17.5.  
Both results indicate that the central 8\hmpc\ of the Shapley supercluster is likely to collapse into 
a more massive system. 
In fact, the density ratios are higher than required for a turn-around, which now implies that the very 
central region of Shapley supercluster has already started to collapse. 
To evaluate the total mass of Shapley with the REFLEX II clusters, we also considered 
the total cluster mass as a function of a radius for the same volume. 
We converted the total cluster mass to the total supercluster mass by adopting a scaling relation found with 
cosmological N-body simulations in~\cite{chon14}.  
We find that the required density ratio for collapse is satisfied out to about 12.4~\hmpc. 
The derived total mass of Shapley within this distance is $1.34\times10^{16}$~\hmass.  
In fact, the turn-around density ratio is already reached at 11.1~\hmpc, which is consistent with
the density ratios estimated from previous work. 
We also note that the filamentary X-ray emission shown in Fig. 2 of~\cite{kull99} coincides with the core of
an already collapsing part of Shapley. 
The estimates of the density ratios of Shapley therefore provide a consistent picture that the central 
11~\hmpc\ is undergoing a collapse meaning that only a central part of Shapley supercluster will form a supercluster 
in the future even if the outskirts of Shapley are also rich in clusters.

\section{Discussion and summary}

In this paper we have emphasised the need for a clearer, more physically motivated definition of superclusters. 
We have shown that defining superstes-clusters as those objects that will collapse in the future leads 
to a conservative selection criterion that does not accept all objects that have been called superclusters 
in the literature, but it leads to a more homogeneous class of objects as seen in our previous work on 
supercluster construction.

Our superstes-cluster definition is also interesting for another reason. 
With this definition, we are selecting the most massive virialised objects that will form in the future. 
In Fig.~\ref{fig:radii} we can, for example, identify the most massive structure in the Millennium simulation 
that will form a virialised dark matter halo in the future with the uppermost point in the plot. 
It has a collapse radius, $r_{v}$ of 17.3 \hmpc\ and a corresponding mass of 1.94$\times$10$^{16}$\msol.
This can be compared to the collapsing fraction of the Shapley supercluster estimated in Sec. 3.3, 
which is quite comparable. 
Shapley is the highest mass concentration found in the nearby Universe in a volume that is quite similar 
to that of the Millennium simulation. 

We find it very encouraging that similar mass estimates are obtained for the most massive structure 
in the observation and and in the simulation with our criteria for selecting superclusters.
We have been applying this criterion in the construction of superstes-clusters using a fof algorithm 
with a linking length tuned to select overdense regions that are close to collapse in the future. 
In our previous study~\citep{chon14} we found a close correspondence of the results of this method with 
the desired overdensity. 
It is, however, not one-to-one, and in particular, we find outliers for very large supercluster sizes, 
which are selected by the fof algorithm but do not reach the overdensity threshold. 
In these extreme cases, superclusters appear as rather elongated filements, and the region that is bound 
to collapse is overestimated by the fof-based method. 

We studied the application of our definition to the Local supercluster and Shapley 
supercluster, as well as Laniakea.
We find that the first two superclusters will collapse in the central regions 
while their outskirts are not gravitationally bound. 
For Laniakea, we find that this structure does not constitute a region with a significant 
overdensity and thus it cannot collapse as a whole. 
While the velocity structure described by~\cite{tully14} highlights an impressively  
large structure in the local Universe, we feel that its labelling as a supercluster is not 
appropriate given that the region is not even overdense. 
It is interesting that \cite{tully14} and other surveys find a local Hubble parameter in 
this region that tends to be higher than the Hubble parameter measured on a more global 
scale~\citep{planckcosm,wmap9}. 
This can be taken as an indication that this region in the local Universe may be rather under-dense. 
This conclusion is supported by our recent studies with X-ray galaxy clusters and studies
of the galaxy distribution (see~\cite{r2density} and~\cite{whitbourn14}).

\begin{acknowledgements}
  We thank the referee for the useful comments. 
  HB acknowledges support from the DfG Transregio Programme TR33 and the DFG cluster of excellence 
  ``Origin and Structure of the Universe'' (www.universe-cluster.de). 
  GC acknowledges the support from the Deutsches Zentrum f\"ur Luft- und Raumfahrt (DLR) with 
  the programme 50 OR 1403. 
  SZ would like to acknowledge the support of The Netherlands Organisation for Scientific Research (NWO) 
  VICI grant and Munich Institute for Astro- and Particle Physics (MIAPP) of the DFG cluster of excellence. 
\end{acknowledgements}


\footnotesize{
  \bibliographystyle{aa}
  \bibliography{scl} 
}

\clearpage

\Online
\begin{appendix}
\section{Appendix}
\begin{table}[h]
  \centering
  \caption{Present-epoch parameters characterising marginally collapsing 
    objects in comparison to those at turn-around for various 
    cosmological models. See text for an explanation of the listed parameters.}
  \begin{tabular}{ccc|ccc|cc}
    \hline
    label & $\Omega_m$ & $h$& $R$ & H$_{\rm ov}$ & $\Delta_c$ & $R$ & $\Delta_c$ \\
    &&&future collapse&&& at turn-around  \\
    \hline
    -&0.24    &  0.7  &  9.811  &  17.60  &  1.36  &   14.31  &  2.43 \\
    -&0.27    &  0.7  &  8.732  &  19.09  &  1.36  &   13.12  &  2.54 \\
    -&0.30    &  0.7  &  7.858  &  20.55  &  1.36  &   12.15  &  2.65 \\
    -&0.33    &  0.7  &  7.134  &  22.00  &  1.35  &   11.35  &  2.75 \\
    WMAP\tablefootmark{a}&0.288   &  0.697&  8.188  &  19.89  &  1.36  &   12.52  &  2.60 \\
    PLANCK\tablefootmark{b}&0.318   &  0.670&  7.409  &  20.50  &  1.36  &   11.66  &  2.71 \\
    -&0.27    &  0.60 &  8.732  &  16.36  &  1.36  &   13.12  &  2.54 \\
    EdS&1.0     &  0.7  & -  &- &- &5.54 & 4.54 \\
    open &0.3     &  0.7  &  2.87 & 52.70& & 12.48 &  \\
    \hline
  \end{tabular}
  \tablefoot{
    \tablefoottext{a}{Best fit for the combined model in~\cite{wmap9}}
    \tablefoottext{b}{Best fit for PLANCK and WMAP polarisation data~\citep{planckcosm}}
  }
  \label{tab:condition}
\end{table}
\end{appendix}

\end{document}